\documentclass[aps,prl,twocolumn,superscriptaddress]{revtex4}

\usepackage{t1enc}
\usepackage{amsmath}
\usepackage{graphicx}
\usepackage{epsfig}
\usepackage{psfrag}
\usepackage{times}

\begin{document}

\title{Symmetry breaking in frustrated systems: effective fluctuation spectrum due to coupling effects}

\author{Alejandro Mendoza-Coto}
\affiliation{Department of Theoretical Physics, Physics Faculty, University of
Havana, La Habana, CP 10400, Cuba}
\affiliation{``Henri-Poincar\'e-Group'' of Complex Systems, Physics Faculty, University of
Havana, La Habana, CP 10400, Cuba}

\author{Rogelio D\'\i az-M\'endez}
\affiliation{Nanophysics Group, Department of Physics,
Electric Engineering Faculty, 
CUJAE, Ave 114 final, La Habana, Cuba}
\affiliation{``Henri-Poincar\'e-Group'' of Complex Systems, Physics
Faculty, University of Havana, La Habana, CP 10400, Cuba}

\author{Roberto Mulet}
\affiliation{Department of Theoretical Physics, Physics Faculty, University of
Havana, La Habana, CP 10400, Cuba}
 \affiliation{``Henri-Poincar\'e-Group'' of Complex Systems, Physics Faculty, University of
Havana, La Habana, CP 10400, Cuba}

\author{Lucas Nicolao}
\affiliation{Dipartimento di Fisica, Universit\`a di Roma "La Sapienza", P.le Aldo Moro 2,
00185 Roma, Italy}

\author{Daniel A. Stariolo}
\affiliation{Departamento de F\'{\i}sica, Universidade Federal do Rio Grande do Sul and\\
National Institute of Science and Technology for Complex Systems CP 15051, 91501-970, Porto Alegre, Brasil}

\date{December  2010}

\begin{abstract}
We study the Langevin dynamics of a $d$-dimensional Ginzburg-Landau Hamiltonian with isotropic
long range repulsive interactions. We show that, once the symmetry is broken,  there is a
coupling between the mean value of the local field and its fluctuations, generating an
anisotropic effective fluctuation spectrum. This anisotropy has many interesting
dynamical consequences. In the infinite time limit, static results are recovered 
which can be compared with the well known Brazovskii's transition to a state with modulated
order.
Our study reveals that the modulated solution appears continuously in $d=3$ contrary
to what is found in the usual approach neglecting coupling of modes, while for $d<3$ 
transitions are still discontinuous. Analytical
results for positional and orientational order parameters are also obtained and 
interpreted in the context of recent discussions.
\end{abstract}

\maketitle

\textit{Introduction.}- According to a classic result due to Brazovskii \cite{Br1975}, 
systems in which the
spectrum of fluctuations have a minimum in a shell in reciprocal space at a non zero wave
vector, undergo a first order phase transition driven by fluctuations to a modulated state,
in contrast to the second order transition predicted by mean field theory. This behavior,
typically produced by the competition of isotropic interactions, have been successfully
tested for two-dimensional models in both theoretical~\cite{garel82} and numerical
studies~\cite{CaStTa2004,dm10} even in systems where interactions are not completely isotropic.
Anyway, despite some efforts~\cite{grousson01,GrKrTaVi2002} there have not been such a clear
confirmation for the character of the phase transition in other dimensions. The existence of
a nematic phase~\cite{abanov95,scannas06,barci07} have also opened the discussion about the  nature of
such a state with orientational but not translational order, that appears above the
Brazovskii transition. Since the strong degeneracy of the fluctuation spectrum gives rise to
the existence of many metastable structures, dynamical effects become very
important~\cite{garel82,GrKrTaVi2002} and a great effort have been
done~\cite{mulet07,TaCo2007,mendoza} in describing equilibrium properties by means of dynamical
parameters. Following that way, in the present work we address some of the above points by
solving the Langevin dynamics of a standard model with competing isotropic interactions,
breaking the symmetry by means of a small external field that is finally turned off. Taking
the infinite time limit, we are able to study the steady state and recover the phase
transition to modulated structures corresponding to that encountered in Brazovskii's
equilibrium calculations~\cite{GrKrTaVi2002,TaCo2007}. Moreover, we show that once the
symmetry is broken, considering non-homogeneous spatial contributions of the average
order parameter on the fluctuations field, leads to the appearance of an effective
fluctuation spectrum that is not isotropic anymore, modifying the usual interpretation
of the low temperature phases. As a result,
for $d=3$ we obtain a continuous transition to the modulated phase, contrary to what is
generally assumed, while the standard Brazovskii phenomenology is recovered when neglecting the
coupling of modes. We find evidence of the existence of the
nematic phase by evaluating analytical expressions for topological order parameters.
This phase is interpreted in terms of the penetration depth of the fluctuations into the
stripes structures, and is contrasted with a novel state also predicted, that consists in the presence
of translational order without the orientational one.

\textit{Model and Formalism.}-
We consider an effective Ginzburg-Landau Hamiltonian
\begin{eqnarray}
 {\cal H}[\phi]  &=& \int d^d x \left[ \frac{1}{2} (\nabla \phi(\vec{x}))^2
+\frac{r}{2}\phi^2(\vec{x})+\frac{u}{4}\phi^4(\vec{x}) \right.\\ \nonumber
&-&\left.\frac{}{} h(\vec{x})\phi(\vec{x}) \right]
+ \frac{1}{2\delta} \int d^d x\,d^d x'\ \phi(\vec{x})\,J(\vec{x},\vec{x}')\,\phi(\vec{x}')
\label{hamilt}
\end{eqnarray}
where $r<0$, $u>0$ and $J(\vec{x},\vec{x}')=J(|\vec{x}-\vec{x}'|)$ represents a repulsive,
isotropic, long range interaction. The external field is represented by $h(\vec{x})$ and the
parameter $\delta$ measures the relative strength between the attractive and repulsive parts
of the Hamiltonian~\cite{mulet07}. The Langevin dynamics takes the form:
\begin{eqnarray}
\nonumber \frac{1}{\Gamma}\frac{\partial \phi(\vec{x},t)}{\partial t}&=& \nabla^2
\phi(\vec{x},t) -r \phi(\vec{x},t)-
u\,\phi^3(x,t)  + h(\vec{x})\\
&-& \frac{1}{\delta}\int d^d
x'\,J(\vec{x},\vec{x}')\phi(\vec{x}',t)+\frac{1}{\Gamma}\eta(\vec{x},t), \label{eqts}
\end{eqnarray}
where $\Gamma$ is set to $1$ for simplicity, $\eta(\vec{x},t)$ represent a Gaussian noise
that usually models the presence of temperature and $h(\vec{x})=h\ \mathrm{cos}(\vec{k}_0
\cdot \vec{x})$ represents a fictitious field that finally tends to zero and is introduced to
break the symmetry. We may write the local field in terms of its fluctuations in the way
$\phi({\vec{x},t})=\left\langle\phi\right\rangle(\vec{x},t)+\Psi({\vec{x},t})$. Averaging
over the thermal noise and initial conditions, the equation of motion (\ref{eqts}) can be
rewritten in terms of the average local field
\begin{eqnarray}
\frac{\partial \left\langle\phi\right\rangle}{\partial t}&=& \nabla^2
\left\langle\phi\right\rangle -r
\left\langle\phi\right\rangle-u\left\langle\phi\right\rangle^3
-3u\left\langle\phi\right\rangle\left\langle \Psi^2\right\rangle
\label{eqs2}
\\
\nonumber
&-& \frac{1}{\delta}\int d^d
x'\,J(\vec{x},\vec{x}')\left\langle\phi\right\rangle(\vec{x}',t)
+h\ \mathrm{cos}(\vec{k}_0 \cdot \vec{x})
\end{eqnarray}
where we have omitted for simplicity the dependence of magnitudes with $(\vec{x},t)$ wherever
it is possible. In the same way we can write from (\ref{eqts}) and (\ref{eqs2}) an equation
for the fluctuations in the form
\begin{eqnarray}
\nonumber
\frac{\partial \Psi}{\partial t}&=& \nabla^2 \Psi -r \Psi-
3u\left\langle \Psi^2\right\rangle \Psi -3u\left\langle\phi\right\rangle^2\Psi\\
&-& \frac{1}{\delta}\int d^d x'\,J(\vec{x},\vec{x}')\Psi(\vec{x}',t)+\eta(\vec{x},t)
\label{eqs4}
\end{eqnarray}
Here we have used the self-consistent Hartree approach, that is,  the substitution of the
term $\langle\,\Psi^3\rangle$ by $3\left\langle\,\Psi^2\right\rangle\,\Psi$ in order to
linearize the final equation.
At this point the problem is to solve equations (\ref{eqs2})
and (\ref{eqs4}) for disordered initial conditions. It is worth to note that, from the
linearity of equation (\ref{eqs4}) and the isotropy of the initial conditions, the quantity
$\langle\Psi^2\rangle$ must be independent of $\vec{x}$.

\textit{Stationary state.}- Now, in the unimodal approximation, we assume
$\langle\phi(\vec{x},t\rightarrow\infty)\rangle=m\ \mathrm{cos}(\vec{k}_0\cdot\vec{x})$ where
$\vec{k}_0$ is the wave vector that minimizes the fluctuation spectrum. From equation (\ref{eqs2}) it is
direct to obtain the following result for $m$
\begin{equation}
m=\sqrt{\frac{4}{3u}\left[-A(k_0)-3u\langle \Psi^2\rangle(\infty)\right]}
\label{m}
\end{equation}
where $A(k_0)=k_0^2+r+\left(1/\delta\right)\hat{J}(k_0)<0$, and the limit $h\rightarrow0^+$
was made.
 Since $\langle\Psi^2\rangle$ is an increasing function of temperature, equality
(\ref{m}) reveals the existence of a dynamical critical temperature $T_c$ where real solutions
cease to exist. The nature of this transition
is not obvious because  $\langle\Psi^2\rangle$ must be obtained solving the whole set of
equations, which may have no solution before $m$ take the zero value.
Now we have to solve
equation (\ref{eqs4}) in the infinite time limit; in the inverse space it can be rewritten
\begin{eqnarray}
\label{eqflu}
\frac{\partial \Psi_{\vec{k}}}{\partial t}&=&\left(A(k_0)+3u\left\langle \Psi^2\right\rangle\right)\Psi_{\vec{k}}-\hat{A}( \vec{k})\Psi_{\vec{k}} \\ 
&+&\left(A(k_0)+3u\left\langle \Psi^2\right\rangle\right)\left(\Psi_{\vec{k}+2\vec{k_0}} + \Psi_{\vec{k}-2\vec{k_0}}\right)
+\eta_{\vec{k}}(t), \nonumber 
\end{eqnarray}
where $\hat{A}(\vec{k})=A(\vec{k})-A(k_0)$. In order to solve the equation system
(\ref{eqflu}) we find the response function. Since there is a coupling between modes having a
difference of momentum $2 \vec{k}_0$, all modes with a difference $2n\vec{k}_0$ are actually
coupled, this makes the exact solution to be a hard task. Instead, what we do is to consider
the truncated system given by the two highest elements of the complete system's matrix. This
approximation captures the essential behavior of the highest eigenvalues of the system as
functions of the momentum $\vec{k}$. These eigenvalues are the responsible of the long time
behavior of the response function. We consider
the highest eigenvalue that leads to a response function that does not diverge in the
infinite time limit. Once we have the form of this response function the calculation of
$\langle\Psi^2\rangle$ can be done. The value of $\langle\Psi^2\rangle$ depends on the
topology of this eigenvalue as a function of $\vec{k}$ in the vicinity of $\vec{k}_0$. 
We use the form $\hat{A}(k)=\frac{A_2}{2}(k-k_0)^2$ usually considered in similar contexts~\cite{mulet07}.
In the infinite time limit we get:
\begin{eqnarray}
\left\langle \Psi^2\right\rangle&=&\frac{4T}{(2\pi)^d}\int_{-k_0}^{k_0}dk_1\int
d^{d-1}k \nonumber 
\\
& & \frac{1}{4\gamma+A_2k_1^2+\frac{A_2}{4k_0^2}\left( \sum_{i>1}^d k_i^2\right)^2},
\label{I2p}
\end{eqnarray}
where $\gamma=-A(k_0)-3u\left\langle \Psi^2\right\rangle$. The expression (\ref{I2p}) is
the main result of this paper. If we compare now with the usual kind of
expressions obtained for $\left\langle \Psi^2\right\rangle$ we see that the system evolve
like if it has an effective fluctuation spectrum of the form
\begin{equation}
A_{\mathrm{eff}}(\vec{k})\propto A_2(k_1-k_0)^2+\frac{A_2}{4k_0^2}\left( \sum_{i>1}^d
k_i^2\right)^2.
\end{equation}
This determine a number of interesting features. The most obvious is the non-equivalence
between spatial directions, this means that the properties of the system are not the same
along the stripes and perpendicular to them. This particular form of the dispersion relation
also has important consequences on the way in which transitions to the modulated phase occur for
different dimensions.

\begin{figure}[!htb]
\includegraphics[width=9cm,height=6cm, angle=0]{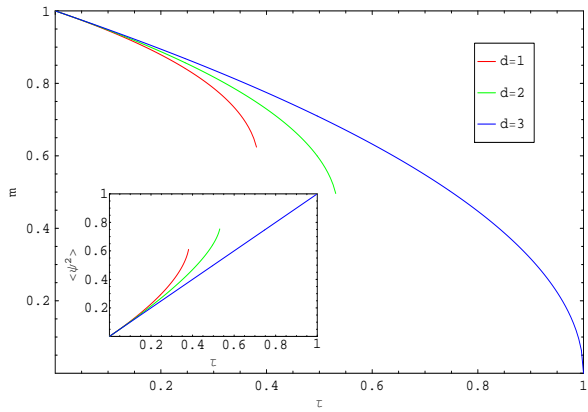}
\caption{Temperature dependence of the amplitude of the principal mode $m$ for different
dimensions. The inset shows the corresponding values of the mean square fluctuations
$\left\langle\psi^2\right\rangle$.}
\label{psiym}
\end{figure}

\textit{Mean square fluctuations.}- In the context of the unimodal approximation $\gamma$
must be a small parameter, even more if we are near the transition. So, in order to study
this transition we may expand integrals around $\gamma=0$. Defining the dimensionless
groups:
\begin{eqnarray}
\left\langle \psi^2\right\rangle=\frac{3u}{-A(k_0)} \left\langle \Psi^2\right\rangle ,\ \ \ \
\tau=\frac{3u\ C(d)}{\left(-A(k_o) \right)^{\frac{3-d}{4}+1} }T, \label{varred}
\end{eqnarray}
where $C(d)$ is an un-interesting constant depending on system dimension, we may obtain after
some calculations
\begin{equation}
\left\langle \psi^2\right\rangle=\frac{\tau}{\left( 1-\left\langle \psi^2\right\rangle\right)^{\frac{3-d}{4}} }.
\label{g2}
\end{equation}
The solution of this equation gives us the value of $\langle\Psi^2\rangle$ for every
temperature and, consequently, the critical temperature below which modulated structures
appears. Contrary to the case $d=3$ where the amplitude of the modulation goes to zero in a
continuous way, for $d=1$ and $d=2$, before the annulation, this amplitude stops having a
real solution. Solving equation (\ref{g2}) for each dimension leads to the following
critical values:
\begin{eqnarray}
\nonumber
d=1, \ \ \ \ \ \ &\tau_c&=\sqrt{\frac{4}{27}},  \ \ \ \ \ \ \left\langle
\psi^2\right\rangle(\tau_c)=\frac{2}{3} \\ \nonumber
d=2, \ \ \ \ \ \ &\tau_c&=0.534,  \ \ \ \ \ \ \left\langle \psi^2\right\rangle(\tau_c)=0.777\\
d=3, \ \ \ \ \ \ &\tau_c&=1,  \ \ \ \ \ \ \ \ \ \ \ \ \ \left\langle
\psi^2\right\rangle(\tau_c)=1,
\end{eqnarray}
as can be seen in figure \ref{psiym}. There the behavior of
$\left\langle\psi^2\right\rangle$ is shown as a function of reduced temperature for every
dimension.
At this point we have obtained self-consistently the fluctuations for subcritical
temperatures from a purely dynamical perspective. We may now analyze the nature of the
transition to the modulated phase. From expression (\ref{m}) it is possible to write the
amplitude of the modulation $m$ as a function of $\langle\psi^2\rangle$ and thus to obtain
the value of $m$ for every temperature in any dimension. 
It is clear from figure \ref{psiym} that, for $d\leq2$, the transition to the
high-temperature paramagnetic phase is discontinuous, contrary to $d=3$, where $m$ goes
continuously to zero as $\tau$ approaches to $\tau_c$. To better realize the
consequences of the coupling between modes in our work let's consider the
dynamic calculation in the infinite time limit without the coupling terms in equation
(\ref{eqflu}). In this case the dynamical equation is diagonal and modes with different
momentum are not coupled. Following standard procedures, the
self-consistent solution to the problem leads to the following equation:
\begin{equation}
\left\langle \Psi^2\right\rangle=\frac{T}{(2\pi)^d}\int d^dk \ \frac{1}{\gamma+\hat{A}(k)},
\end{equation}
where $\gamma=-A(k_0)-3u\left\langle \Psi^2\right\rangle$ as before. What we have obtained is
a new equation for the square mean fluctuations in which the fluctuation spectrum is
spherically symmetric as in the standard Brazovskii-like approaches. The solution to this
problem has been obtained many times in the literature~\cite{GrKrTaVi2002,TaCo2007} showing
discontinuous transitions to the striped phase. Our new results suggest that the nature
of the transition to the modulated phase depends on dimensionality.

\begin{figure}[!htb]
\includegraphics[width=9cm,height=6cm,angle=0]{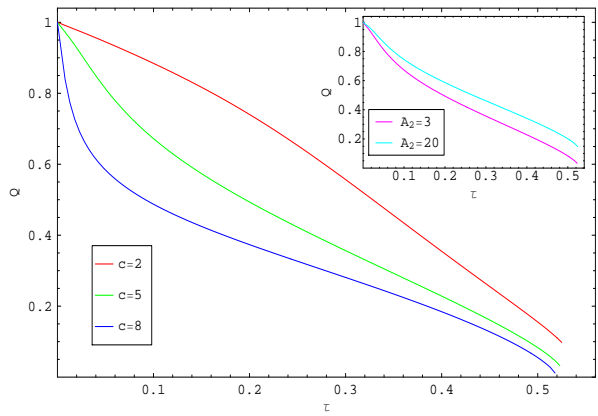}
\caption{Behavior of $Q$ as a function of the reduced temperature. The main figure
corresponds to $A_2=3$ and several values of $c$. In the inset $c=5$ and several values of
$A_2$ were taken.}
\label{Q}
\end{figure}

\textit{Topological properties.}- We now focus in the two-dimensional case in order to study
the topological order of the modulated phase by means of two parameters commonly used in the
literature with this purpose.\cite{nicolao07,quian03} The first one is a positional
order parameter defined by:
\begin{equation}
M=\left\langle \frac{1}{S}\int d^2x \ \phi(\vec{x})\
\mathrm{sin}(\vec{k}_0\cdot\vec{x})\right\rangle, \label{trp}
\end{equation}
where $S$ is the area of the system. As can be seen, this parameter corresponds to a
staggered normalized magnetization which is $1$ only at zero temperature. 
Orientational order of the stripes can be quantified by means of the parameter
\begin{equation}
Q=\left\langle\int d^2x\ \mathrm{cos}(2\theta(\vec{x}))\ \delta(\phi(\vec{x}))\right\rangle
/\left\langle\int d^2x\ \delta(\phi(\vec{x}))\right\rangle \label{orp}
\end{equation}
which again is defined in such a way that is $1$ only in the perfect stripes configuration.
Here $\delta(\phi(\vec{x})) = \phi(\vec x + \delta \vec x)-\phi(\vec x)$.
\begin{figure}[!htb]
\includegraphics[width=9cm,height=6cm,angle=0]{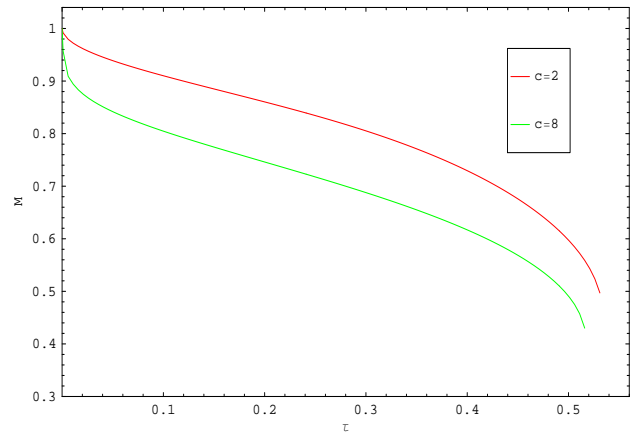}
\caption{Behavior of $M$ as a function of the reduced temperature for different values of
$c$} \label{figP}
\end{figure}
In order to calculate parameters (\ref{trp}) and (\ref{orp}) we use a gaussian distribution
for $\phi$ and its spatial derivatives in the evaluation of the corresponding averages. This
assumption is consistent with the Hartree approximation and allows us to find analytically
closed expressions that, as far as we know, have never been obtained for topological order
parameters. Since the final expressions are not particularly enlightning, we numerically
evaluate them under some reasonable approximations. 
In this context we observe
that the positional order parameter only depends on the dimensionless quantity
$c=\frac{-A(k_0)}{k_0^2}$. In figure \ref{figP} we show typical behaviors of this
positional parameter. On the other hand, the orientational order parameter  depends on $c$
and also on the number $A_2$. This dependence is shown in figure \ref{Q}. As can be expected,
the presence of a critical temperature for the two-dimensional system is reflected by means
of jumps to zero of both, orientational and positional parameters at $T_c$ (figures
\ref{figP} and \ref{Q}). Our calculations show that these parameters decrease with
increasing $c$, while the increase of $A_2$ only causes an increasing of the orientational
order. This dependence of order parameters with $A_2$ and $c$ reveals a rich topological
scenario for the two-dimensional system. In particular, this picture supports the existence
of the so-called nematic phase that has been the subject of some debate in the recent
literature.\cite{nicolao07,barci07,scannas06} This nematic phase is characterized by a strong
orientational order in the absence of positional order. Our results suggests that, by
appropriately increasing $c$ and $A_2$, we may eventually obtain a region of temperatures close
to $T_c$ in which the orientational order prevails over the positional one. To test this one
should have to consider a particular model and its full fluctuation spectrum.
On the other hand, the existence of a phase in which positional order prevails over
orientational one is also suggested. From figure \ref{Q} one can see that increasing $c$
the decay of the orientational parameter at low temperatures becomes more abrupt. This fact,
together with the decrease of $Q$ for increasing values of $A_2$ (inset of figure \ref{Q})
could be responsible for the appearance of such a phase. One can visualize this state by
considering highly irregular stripe borders confined in such a small width that the staggered
magnetization $M$ is not too compromised. In that sense the lack of positional order in
the nematic phase could be seen as the situation in which this width, that defines the
penetration depth of the fluctuations, completely covers the width of the stripes. In this
situation, thermal fluctuations are strong enough to break the stripes order generating a
configuration with topological defects that, as have been discussed in previous works, may
give rise to a nematic phase. Finally, there is an absence of numerical evidence in
literature supporting the existence of a phase with positional but not orientational
order. This is probably due to the presence of a strong anisotropy, caused by a
discretization mesh of the order of the modulation length in numerical
works~\cite{nicolao07,scannas06}. In fact, this anisotropy effects favors the emergence of 
orientational order that makes the nematic phase easier to find.

\textit{Conclusions.}- Taking the infinite time limit, the steady state properties of a model
of competing interactions have been studied from a purely dynamical calculation. Such a
procedure makes use of the Hartree approximation over the fluctuations field and 
the isotropy of the system is explicitly broken. This leads to an effective fluctuation spectrum that is not
isotropic anymore, contrary to the one present in Brazovskii-like calculations. Within this
formalism, it was possible to analytically obtain the fluctuations as functions of the
temperature and thus analyze some interesting parameters in the modulated state. In
particular, the first order character suggested in literature for this
transition is not verified for the tree-dimensional case, in which a continuous transition
takes place due to the coupling between modes. The present approach 
extends our understanding about the nature of modulated phases in systems with competing
interactions. We have also calculated analytically and discussed usual
topological observables. Our restuls suggest that, for
certain range of temperatures close to the transition, it may exist a nematic phase with purely
orientational order, or even a
phase in which the translational order prevails over the orientational one, depending on
some system dependent parameters. A more detailed
paper including the discussion of dynamical observables will be published in short. 

We acknowledge useful discussions with D. G. Barci at the early stages of this work.
D.A.S. acknowledges partial support from CNPq (Brazil).

\bibliography{paper_sb_prl_v1}

\end{document}